\documentclass[12pt]{article}

\usepackage[cp1251]{inputenc}  %для WiN
\usepackage[T2A, OT1]{fontenc}
\usepackage[english]{babel}
\usepackage{amsfonts, longtable}
\usepackage{amsmath, amssymb, amsthm}
\usepackage{cite}
\usepackage{indentfirst}

\newtheorem{theo}{Theorem}%[section]
\newtheorem{predl}[theo]{Statement}%[section]
\newtheorem{lem}[theo]{Lemma}%[section]
\newtheorem{sld}[theo]{Corollary}%[theo]
\newtheorem{prop}[theo]{Property}%[chapter]

\newcommand{\comment}[1]{}

\newenvironment{prf}{\par\textbf{Proof.}}{\par\qed\par}

\def\Ls{{{\mathcal L}}}

\title{Some Closed Classes of Three-Valued Logic Generated by Periodic Symmetric Functions}
\author{A.\,V.\,Mikhailovich\footnote{National Research University Higher School of Economics}}
\date{April, 2016}
\begin{document}
	\maketitle
	\begin{abstract}
		Closed classes of three-valued logic generated by periodic symmetric funtions that equal $1$ in tuples from $\{1,2\}^n$ and equal $0$ on the rest tuples are considered. Criteria for bases existence and finite bases existence for these classes is obtained.
	\end{abstract}
	The problem of the bases existence for some families of closed 
	classes of the three-valued logic functions
	is considered in the paper. 
	E.\,Post~\cite{post41} (see also, for instance,~\cite{lau06}) described all closed classes of Boolean functions and showed that each
	such class has a finite basis. This result is not extendable to the case of $k$-valued logics for 
	$k\geq 3.$ Ju.\,I.\,Janov and A.\,A.\,Muchnik~\cite{janov} (see also, for instance,~\cite{lau06}) 
	showed that for all $k\geq 3$ the set $P_k$ (here $P_k$ is the set of all functions of the $k$-valued logic) contains closed classes having a countable basis, and those having no basis. The generating
	systems for classes from these examples consist of symmetric functions 
	that take values from the set $\{0, 1\}$
	and equal to zero on tuples containing at least one zero component.
	Similar classes have been described in~\cite{mikh08, mikh09, mikh12}. Criteria 
	of basis existence and finite basis existence for these classes have been obtained.
	In~\cite{mikh13} some closed classes, generated by symmetric periodic function
	with bounded period have been investigated. Criterium of finite basis existence has
	been obtain and it was shown that if such class has no finite basis, it has no any basis. This paper deals with closed classes of three-valued logic, generated by symmetric periodic functions with increasing period. Criteria of basis existence and 
	finite basis existence has been obtained.
	
	Let $R$ be the set of all functions that take values from the set $\{0,1\}$
	and equal zero on tuples containing at least one zero\footnote{All necessary definitions can be found in~\cite{lau06,mikh08,mikh09,mikh12}.}. 
	\begin{multline*}
	R=\{f(x_1,\ldots,x_n)\mid ((\forall\widetilde\alpha)((\widetilde\alpha\in\{0,1,2\}^n)
	\rightarrow (f(\widetilde\alpha)\in\{0,1\}))) 
	\& \\((\forall\widetilde\alpha)((\widetilde\alpha\in\{0,1,2\}^n\backslash\{1,2\}^n)
	\rightarrow (f(\widetilde\alpha)=0)))\}
	\end{multline*}
	In this paper we deal with some subclasses of the class $R$.
	Any function that does not change with variable relabeling is called symmetric.
	Denote by $S$ the set of all symmetric functions from $R$.
	The set of all tuples that can be obtained from each other by 
	component permutation is called a layer. 
	With $\mathcal L(e,d)$ we denote a layer from $\{1,2\}^n$
	containing $e$ $1$s and $d$ $2$s.
	Let $N_f=\{\widetilde\alpha\in E_3^n\mid f(\widetilde\alpha)=1\}.$
	With $i_s(x_1,\ldots,x_s)$ we denote a function from $R$ 
	such that $N_{i_s}=\{1,2\}^s.$
	Let
	$$
	I=\bigcup_{n=1}^{\infty} \{i_n\}.
	$$
	
	Let $\Phi$ be a formula over $R,$ $\Phi_1$ be subformula of the formula $\Phi$,
	$\Phi_1$ realize function $f(x_1,\ldots,x_n).$ Subformula $\Phi_1$ is called 
	essential if $f(x_1,\ldots, x_n)$ is not equivalent to $i_n(x_1,\ldots,x_n).$
	
	Let $\Phi$ be formula over $R.$ Denote by $\Theta(\Phi)$ the set of all functions 
	$g(x_1,\ldots,x_m)$ that satisfies the following conditions:
	\begin{enumerate}
		\item[$1.$] There exist subformula $\Phi_1$ of the formula $\Phi$ that has the 
		form $g(\mathcal B_1,\ldots, \mathcal B_m)$, where 
		$\mathcal B_1,\ldots, \mathcal B_m$ are formulas over $R.$
		\item[$2.$] Replacement the subformula $\Phi_1$ by the formula 
		$i_m(\mathcal B_1,\ldots, \mathcal B_m)$ in the formula $\Phi$ gives a
		formula that is not equivalent to the formula $\Phi.$
	\end{enumerate}
		
	A function $f,$ $f\in S$ is called periodic with period equal $t$ if there exist $e_f,$ $d_f,$
	$e_f+d_f=n,$ $0\leq d_f<t$ such that 
	$$
	N_f=\bigcup_{i=0}^s \mathcal L(e_f-it, d_f+it), \mbox{ where } s=\left]\frac{n-d_f}{t}\right[.
	$$
	Denote by $t_f$ period of the function $f.$
	Denote by $PS$ the set of all periodic symmetric non-zero functions.
	Denote by $PS^t$ the set of all periodic symmetric functions with period $t.$
	Let 
	$$
	PS^{(t)}=\bigcup_{i\leq t} PS^{i}.
	$$
	Let $p\in\mathbb N,$ $p$ is prime number. Then let
	$$
	PS^{[p]}=\bigcup_{i\leq t} PS^{p^i}.
	$$
	Let $p_1,\ldots, p_s\in\mathbb N,$ $p_1,\ldots,p_s$ are prime numbers. Then let
	$$
	PS^{[p_1,p_2,\ldots,p_s]}=\bigcup_{j_1\geq 0,\ldots, j_s\geq 0} PS^{p_1^{j_1}\cdot\ldots\cdot p_s^{j_s}}.
	$$

%\begin{prop}
%	Let $f(x_1,\ldots,x_n),$ $g(x_1,\ldots,x_n)\in PS,$ $h(x_1,\ldots,x_n)\in R,$
%	$h\not\equiv 0,$ $N_h=N_f\cap N_g.$ Then $h\in PS$ and $t_h=lcm(t_f, t_g).$
%\end{prop}
	
\begin{prop}
	Let $f_1(x_1,\ldots,x_n),$ $f_2(x_1,\ldots,x_n),$ \ldots, $f_s(x_1,\ldots,x_n)\in PS,$
	$d_{f_1}=d_{f_2}=\ldots=$ $d_{f_s}=0,$ $h(x_1,\ldots,x_n)\in R,$
	$h\not\equiv 0,$ $N_h=N_{f_1}\cap N_{f_2}\cap\ldots\cap N_{f_s}.$ Then $h\in PS$ and $t_h=lcm(t_{f_1},t_{f_2},\ldots,t_{f_s}).$
	\label{sim_per_nf_intersec}
\end{prop}

\begin{prop}
	Let $l,m,n\in \mathbb N,$ $n\geq 2.$ Then
	\begin{enumerate}
		\item[$1.$] $i_l(i_m(x_1,\ldots,x_m),x_{m+1},\ldots,x_{l+m-1})=
		i_{l+m-1}(x_1,\ldots,x_{l+m-1}).$
		\item[$2.$] $i_n(x_1,\ldots,x_{n-1},x_{n-1})=i_{n-1}(x_1,\ldots,x_{n-1}).$
		\item[$3.$] $i_m\in[\{i_n\}].$
		\item[$4.$] $I=[I].$
		\item[$5.$] $I=[\{i_n\}].$
	\end{enumerate}
\end{prop}
	
	\begin{predl}
		Let $\Phi$ be a formula over $R;$
		$\Phi_1$ be a subformula of $\Phi.$
		Then for any tuple $\widetilde\alpha$ 
		the equality $\Phi_1(\widetilde\alpha)=0$ implies
		the equality $\Phi(\widetilde\alpha)=0.$
		\label{mikh_mos_lemma10}
	\end{predl}
	\begin{prf}
		Suppose $d$ is the depth of $\Phi,$ $d_1$ is the depth of $\Phi_1.$
		Let $\Phi$ have the form % после let всегда infinitive
		$g(\mathcal B_1,\ldots,\mathcal B_m).$ 
		
		The proof is by induction on $d-d_1.$
		For $d-d_1=0,$ the proof is trivial.
		Let $d-d_1=1.$ 
		Then $\Phi_1=\mathcal B_t,$ $1\leq t\leq m.$
		Since $g\in R,$ 
		$\mathcal B_t(\widetilde\alpha)=0$ implies $\Phi(\widetilde\alpha)=0.$
		
		Let $d-d_1>1.$
		%Let $\Phi$ has the form 
		%$g(\mathcal B_1,\ldots,\mathcal B_m),$ $\widetilde\alpha\in E_3^n.$ 
		Then $\Phi_1$ is a subformula of $\mathcal B_t$ for $1\leq t\leq m.$
		Denote by $d_2$ the depth of $\mathcal B_t.$
		Then $d_2-d_1<d-d_1.$
		By the inductive assumtion $\mathcal B_t(\widetilde\alpha)=0.$
		Note that $d-d_2<d-d_1.$
		Then by the inductive assumtion $\Phi(\widetilde\alpha)=0.$
	\end{prf}
	
	\begin{sld}
		If a formula over $R$ equals $1$ on a tuple$,$ 
		then any subformula of the formula also equals $1$ on the same tuple.
		\label{mikh_mos_sld1010}
	\end{sld}
	
	\begin{sld}
		If $\Phi$ is a formula over $R$ and 
		$\Phi_1$ is a subformula of formula $\Phi$ then $N_{\Phi}\subset N_{\Phi_1}.$
		\label{mikh_mos_sld1020}
	\end{sld}

\begin{predl}
	Let $f(x_1,\ldots, x_n)\in PS^t,$ $t>1,$ there exist tuple $\widetilde{\alpha}\in N_f$
	such that $0<|\widetilde{\alpha}|<n.$ Let $\Phi$ be formula over $R$ that realize 
	function $f$, $\Phi_1$ be subformula of the formula $\Phi$, $\Phi_1$
	has the form $g(\mathcal B_1,\ldots, \mathcal B_m),$ where $g\in PS^r,$
	$r\in \mathbb N,$ $\mathcal B_1,\ldots, \mathcal B_m$ are formulas over $R$.
	Suppose that among the formulas $\mathcal B_1,\ldots, \mathcal B_m$ there are
	$q_1$ symbols of variable $x_1,$ $q_2$ symbols of varialbe $x_2,$ \ldots,
	$q_n$ symbols of variable $x_n.$ Then for any $i, j,$ $1\leq i, j\leq n$ there exists
	$l\in \mathbb Z^+$ such that $q_i-q_j=lr.$
	\label{sim_per_numofvar}
\end{predl}	

The proof is based on Corollary~\ref{mikh_mos_sld1020} and the fact that function
does not change with variables relabeling.
	
\begin{sld}
	Let $f(x_1,\ldots, x_n)\in PS^t,$ $t>1,$ $\Phi$ be formula over R that realize function $f,$ $\Phi_1$ be subformula of the formula $\Phi$, $\Phi_1$ has the form
	$g(\mathcal B_1,\ldots, \mathcal B_m),$ where $g\in PS^r,$
	$r\in \mathbb N,$ $\mathcal B_1,\ldots, \mathcal B_m$ are formulas over $R$.
	If among the formulas $\mathcal B_1,\ldots, \mathcal B_m$ there are exactly 
	$k_i r$ symbols of the variable $x_i$ for some $i,$ $1\leq i\leq n,$ $k_i,$ 
	$k_i\in\mathbb Z^+,$ then $\Phi_1$ is equivalent to the formula 	
	$i_m(\mathcal B_1,\ldots, \mathcal B_m).$
	\label{sim_per_subfreplace}
\end{sld}

\begin{sld}
	Let $f(x_1,\ldots, x_n)\in PS^t,$ $t>1,$ $\Phi$ be formula over R that realize function $f.$ Then there exists at least one subformula $\Phi_1$ of the formula $\Phi$ that has the form $g(\mathcal B_1,\ldots, \mathcal B_m),$ where $g\in PS^r,$
	$r\in \mathbb N,$ $\mathcal B_1,\ldots, \mathcal B_m$ are formulas over $R$
	and among the formulas $\mathcal B_1,\ldots, \mathcal B_m$ there are
	$q_1$ symbols of variable $x_1,$ $q_2$ symbols of varialbe $x_2,$ \ldots,
	$q_n$ symbols of variable $x_n,$ such that $q_i>0$ and $q_i$ is not
	divisible by $r$ for all $i=1,\ldots,n.$
\end{sld}	

\begin{sld}
		Let $f(x_1,\ldots, x_n)\in PS^t,$ $t>1,$ $\Phi$ be formula over R that realize function $f,$ $\Phi_1$ be essential subformula of the formula $\Phi$, 
		$\Phi_1$ has the form
		$g(\mathcal B_1,\ldots, \mathcal B_m),$ where $g\in PS,$
		$\mathcal B_1,\ldots, \mathcal B_m$ are formulas over $PS$.
		Then for all $i=1,\ldots,n,$ there is symbol of the variable $x_i$  among 
		$\mathcal B_1,\ldots, \mathcal B_m.$		
		\label{sim_per_everyvar}
\end{sld}

\begin{sld}
	Let $f(x_1,\ldots,x_n)\in PS^t,$ $t>1,$ $G\subset PS^t,$ $f\in[G].$ Then there
	exist function $g(x_1,\ldots, x_m)\in G$ such that $m\geq n.$ 
	At that if $m=n$ then $f$ is congruent to $g.$
	\label{sim_per_varnum}
\end{sld}

\begin{predl}
	Let $f(x_1,\ldots,x_n)\in PS^t,$ $t>1$ $\Phi$ be formula over $PS$ that
	realize function $f,$ $\Phi_1$ be essential subformula of the formula $\Phi,$ 
	$\Phi_1$ does not have other essential subformulas, 
	$\Phi_1$ has the form $g(\mathcal B_1,\ldots, \mathcal B_m),$ 
	where $g\in PS^r,$ $r\in \mathbb N,$ $\mathcal B_1,\ldots, \mathcal B_m$ 
	are formulas over $PS$. Let formula $\Phi_1$ realize function 
	$h(x_1,\ldots, x_l).$ Then $h\in PS^w$ for some $w>1$ and $r$ is 
	divisible by $r.$
\end{predl}

The proof follows from Corollary~\ref{mikh_mos_sld1020} and definition of the 
symmetric periodic functions.

\begin{sld}
	Let $f(x_1,\ldots, x_n)\in PS^t,$ $t>1,$ $\Phi$ be formula over $PS$ that realize
	function $f,$ $\Phi_1$ be essential subformula of the formula $\Phi,$  that 
	realize function $h.$ Then $h\in PS^w$ for some $w$ and $t$ is divisible by $w.$
\label{sim_per_esssubform}
\end{sld}

\begin{sld}
	Let $f(x_1,\ldots, x_n)\in PS^{[p]}$ $p>1,$ $t$ be simple number, $\Phi$ be formula over $PS$ that realize function $f,$ $\Phi_1$ be essential subformula of the formula $\Phi,$ that has the form $g(\mathcal B_1,\ldots, \mathcal B_m),$ 
	where $g\in PS^r,$ $r\in \mathbb N,$ $\mathcal B_1,\ldots, \mathcal B_m$ 
	are formulas over $PS.$ Then $\Phi_1$ realize $f$ and $f\in [\{g\}\cup I]$ and
	$t_f=t_g.$
	\label{sim_per_gener}
\end{sld}

\begin{predl}
	For all $r, t\in\mathbb N,$  $f(x_1,\ldots, x_n)\in PS^t$ it holds
	$$
	f\in[PS^r\cup I]\Leftrightarrow r \mbox{ is divisible by } t.
	$$ 
\end{predl}

The proof is based on Corollary~\ref{sim_per_esssubform} and definition of 
the symmetric periodic functions.

\begin{lem}
	Let $f(x_1,\ldots,x_n),$ $g(x_1,\ldots, x_m)\in PS,$ 
	$\{(1^m),(2^m)\}\not\subset N_g,$ $t_f>1,$ $d_f+t_f\leq n.$ Then
	\begin{enumerate}
		\item[$1.$] $f\in [\{g\}]$ iff
		\begin{itemize} 
			\item $\exists t\in\mathbb N$ such that $t=\frac{t_g}{t_f};$
			\item $d_g$ is divisible by $t\cdot (d_f, t_f);$
			\item $m=qn+st_g,$ where $s\in \mathbb Z^+,$ 
			$q\in \mathbb N,$ $0<q<t_g$ such that $d_g+kt_g=qd_f$ 
			for some $k\in\mathbb Z^+.$
			\item $q$ is divisible by $t.$
		\end{itemize}
		\item[$2.$] $f\in [\{g\}\cup I]$ iff
		\begin{itemize} 
			\item $\exists t\in\mathbb N$ such that $t=\frac{t_g}{t_f};$
			\item $d_g$ is divisible by $t\cdot (d_f, t_f);$
			\item $m\geq qn$ where
			$q\in \mathbb N,$ $0<q<t_g$ such that $d_g+kt_g=qd_f$ 
			for some $k\in\mathbb Z^+.$
			\item $q$ is divisible by $t.$
		\end{itemize}
	\end{enumerate}
	\label{sim_per_order}
\end{lem}

The proof is based on Statements~\ref{sim_per_numofvar} and \ref{sim_per_subfreplace}.
	
\begin{lem}
		Let $f(x_1,\ldots,x_n),$ $g(x_1,\ldots, x_m)\in PS,$ 
		$\{(1^m),(2^m)\}\subset N_g,$ $t_f>1.$ Then the following statements are
		equivalent:
		\begin{enumerate}
			\item[$1.$] $f\in[\{g\}];$
			\item[$2.$] $f\in[\{g\}\cup I];$
			\item[$3.$] $d_f=0,$ $\frac{t_g}{t_f}$ is the whole number, $m\geq \frac{t_g}{t_f}\cdot n.$
		\end{enumerate}
\end{lem}

Since $i_2\in[\{g\}]$ the proof follows from Lemma~\ref{sim_per_order}.

\begin{lem}
	Let $G\subseteq PS^{[p]},$ $\forall f(x_1,\ldots,x_n)\in G$ 
	$d_f=0,$ $\exists g(x_1,\ldots,x_m)\in G,$ $(2^m)\in N_g.$ Then the following statements are equivalent:
		\begin{enumerate}
			\item[$1.$] $[G]$ has basis.
			\item[$2.$] $[G]$ has finite basis.
		\end{enumerate}
	\label{sim_per_dfzero}
\end{lem}

To prove the lemma it is enough to show that for every function $f\in G$ 
the set $G_f=$ $\{h\in G\mid h\notin[\{f,g\}] \mbox{ and } f\notin [\{h,g\}]\}$ 
contains finite number of non-congruent functions. Indeed,
Lemma~\ref{sim_per_order} implies that if $f\in [\{h,g\}]\backslash[\{g\}],$ 
then either $n_h>n_f$ and $\frac{n_f}{t_f}>\frac{n_h}{t_h}$
or $n_h<n_f$. Obviously, there are finite number of such functions. Moreover, 
for any function $f\in PS$ the set $PS\cap[\{f\}]$ contains only finite number of non-congruent functions.

\begin{lem}
	Let $G\subseteq PS^{[p]},$ $\forall f(x_1,\ldots,x_n)\in G$ 
	$d_f=0,$ $\exists g(x_1,\ldots,x_m)\in G,$ $(2^m)\in N_g,$
	$G$ is infinite, $G$ not contain congruent functions.
	Let $F\subset PS,$ $G\subset P.$ Then $F$ has no basis.
\label{sim_per_nobasis}
\end{lem}

The proof is similar to the proof of Lemma~\ref{sim_per_dfzero}.

\begin{theo}
	Let $p$ be prime number, $G\subset PS^{[p]},$ $G$ not contain congruent functions, $F=[G].$ Then 
	\begin{itemize}
		\item [$1.$] Class $F$ has a finite basis iff $G\backslash I$ is finite.
		\item [$2.$] Class $F$ has an infinite basis iff $G\backslash I$ is infinite
		and for any $t\in \mathbb N$ the set of all functions $f$ such that
		$\frac{t_f}{(d_f,t_f)}=p^t$ is finite.
		\item [$3.$] Class $F$ has no basis iff there exitst 
		$t\in \mathbb N$ such that set of all functions $f$ such that
		$\frac{t_f}{(d_f,t_f)}=p^t$ is infinite.
	\end{itemize}
\end{theo}

\begin{prf}
	The statement 1 of the Theorem follows from Lemma~\ref{sim_per_varnum}.
	
	Let us prove the statement 2 of the Theorem. Let $F$ have infinite basis $\mathfrak A.$  Let
	$G_0=$ $\{g\in G\mid d_g=0\}.$ It follows from Lemma~\ref{sim_per_nobasis} that
	$G_0$ is finite.
	
	Let $f\in \mathfrak A$ and formula $\Upsilon_f$ realize function $f$ over $G$.
	Let $\Phi$ is arbitrary formula over $A.$ Denote by $\pi(\Phi)$ formula over $G$
	that is obtained from formula $\Phi$ by replacement every function $f$ by 
	corresponding formula $\Upsilon_f.$
	Let $\mathfrak A_0=\{f\in\mathfrak A\mid \exists g\in G_0\cap \Theta (\Upsilon_f)\}.$ It can be proved that $\mathfrak A_0$ is finite. 
	Hence, using Lemma~\ref{sim_per_nobasis} we obtain that 
	$(G\backslash (G\cap[\mathfrak A\backslash\mathfrak A_0]\cup I))$ contains finite number of 
	non-congruent functions.

	Let $I\subset F$ (case $I\not\subset F$ is similar). Let 
	$$
	\mathfrak B=\{g\in G \mid \forall g'\in G (( g\in [\{g'\}\cup I])\Rightarrow 
	(g'\in[\{g\}\cup I]))\}.
	$$
	It is easily shown that for every $s\in\mathbb N$ 
	the set $\mathfrak B\cap PS^{[p^s]}$ is finite.
	
	Consider $f\in \mathfrak A\backslash \mathfrak A_0.$ 
	Since $\mathfrak A$ is basis there
	exist function $g\in \Theta(\Upsilon_f)$ 
	such that $g\notin[\mathfrak A\backslash \{f\}].$
	It is easily shown that formula $\Upsilon_f$
	contains subformula that has the form
	$g(x_{i_1},\ldots,x_{i_m}),$ such that $i_j\neq i_k$ for $j\neq k.$
	If there are at least two non-congruent functions $g_1,$ $g_2$
	such that $g_1(x_1,\ldots, x_n),$ $g_2(x_1,\ldots, x_n)$ $\in \Theta (\Upsilon_f)$ and 
	$g_1, g_2\notin [\mathfrak A\backslash \{f\}],$ then 
	there are formulas $\Phi_1$ and $\Phi_2$ over $\mathfrak A$
	that realize functions $g_1$ and $g_2$ respectively and 
	$g_1\in\Theta(\pi(\Phi_2))$ and $g_2\in\Theta(\pi(\Phi_1)).$
	Statement~\ref{sim_per_varnum} implies that $m=n$. 
	Since $N_{g_1}=N_{g_2}$ and functions $g_1$ and $g_2$ 
	are symmetric, $g_1(x_1,\ldots,x_n)=g_2(x_1,\ldots,x_n).$
	Denote by $g_f$ functions $g\in\Theta(\Upsilon_f)$ such that 
	$g\notin[\mathfrak A\backslash \{f\}].$
	Moreover, all simple subformula of the formula $\Upsilon_f$
	have the form $g(x_{i_1},\ldots,x_{i_m}),$ $i_j\neq i_k$ for all 
	$1\leq i,k\leq m.$ 
	
	If there exist function $g'$ such that $g\in[\{g'\}\cup I],$
	$g'\notin[\{g\}\cup I],$ then $g'\notin [\mathfrak A\backslash \{f\}].$ 
	Indeed, it was shown above that there for every function $f\in\mathfrak A\backslash 
	\mathfrak A_0$ there is only one function $g\in\Theta(\Upsilon_f)$ such that 
	$g\notin [\mathfrak A\backslash\{f\}].$ Hence, $f\in[\mathfrak A\backslash\{f\}\cup \{g'\}]=[\mathfrak A\backslash \{f\}].$ That contradicts definition of basis. 
	Hence, $g\in\mathfrak B.$
	
	Consider $h\in \Theta(\Upsilon_f)\backslash\mathfrak B.$
	Since $g_f\notin \mathfrak A\backslash \{f\},$ formula $\Phi_g$
	that realize function $g_f$ contains subformula of the form
	$f(x_{i_1},\ldots,x_{i_n}).$ Hence, formula $\pi(\Phi_g)$
	contains subformula of the form $h(\mathcal B_1,\ldots, \mathcal B_s)$.
	Statement~\ref{sim_per_numofvar}, Property~\ref{sim_per_nf_intersec} and 
	Lemma~\ref{sim_per_order} implies that $g_f\in [\{h\}\cup I]$.
	That contradicts relation $g_f\in \mathfrak B.$ Hence, $\Theta(\Upsilon_f)\in \mathfrak B.$

	Consider $h\in G\cap[\mathfrak A\backslash\mathfrak A_0]$. 
	Let formula $\Phi_h$ realize function $h$ over $\mathfrak A.$ Consider formula 
	$\pi(\Phi_h).$ Corollary~\ref{sim_per_gener} implies that 
	there is function $g\in\Theta(\pi(\Phi_g))$
	such that $h\in[\{g\}\cup I].$ It can be easily checked that there is function 
	$f\in\Theta(\Phi_h)$ such that $g\in\Theta(\Upsilon_f).$
	Since $\Theta(\Upsilon_f)\in \mathfrak B,$ and the set 
	$(G\backslash (G\cap[\mathfrak A\backslash\mathfrak A_0]\cup I))$ contains finite number of non-congruent functions, for any function $h'\in G$ there exist
	funciton $g'\in \mathfrak B$ such that $h'\in[\{g\}\cup I].$
	
	It can be easily proved that the set $\mathfrak B$ or $\mathfrak B\cup \{i_2\}$ is 
	basis of the class $F$. Since $G\backslash I$ is infinite, the basis is also infinite.
	
	Using Lemma~\ref{sim_per_order} it can be prooved that 
	for any function $h'\in G$ there exist
	funciton $g'\in \mathfrak B$ such that $h'\in[\{g\}\cup I]$ iff for any 
	$t\in\mathbb N$ the set of all function $f$ such that 
	$\frac{t_f}{(d_f, t_f)}=p^t$ is finite. 
	
	The statement 3 follows from statements 1 and 2 of the Theorem.
\end{prf}

\medskip 

This study (research grant No 14-01-0144) supported by The National Research University~--- Higher School of Economics' Academic Fund Program in 2014/2015.

\end{document}